# Direct observation of the quantum-fluctuation driven amplitude mode in a microcavity polariton condensate

Mark Steger[*,1], Ryo Hanai[2,3], Alexander Orson Edelman[2,4], Peter B Littlewood[2,4], David W Snoke[5], Jonathan Beaumariage[5], Brian Fluegel[1], Ken West[6], Loren N. Pfeiffer[6], Angelo Mascarenhas[1]

[1]National Renewable Energy Lab, Golden, CO 80401, USA

[2]James Franck Institute and Department of Physics, University of Chicago, Chicago, Illinois 60637, USA

[3]Department of Physics, Osaka University, Toyonaka 560-0043, Japan

[4]Physical Sciences and Engineering, Argonne National Laboratory, Argonne, Illinois 60439, USA

[5]Department of Physics and Astronomy, University of Pittsburgh, Pittsburgh, PA 15260, USA

[6]Department of Electrical Engineering, Princeton University, Princeton, NJ 08544, USA

*Mark.Steger@NREL.gov

Direct observation of quantum-fluctuation driven amplitude mode in a microcavity polariton condensate

Mark Steger, Ryo Hanai, Alexander Orson Edelman, Peter B Littlewood, David W Snoke, Jonathan Beaumariage,

Brian Fluegel, Ken West, Loren Pfeiffer, Angelo Mascarenhas

**The Higgs amplitude mode is a collective excitation studied and observed in a broad class of matter, including superconductors, charge density waves, antiferromagnets, $^3$He p-wave superfluid, and ultracold atomic condensates. In all the observations reported thus far, the amplitude mode was excited by perturbing the condensate out of equilibrium. Studying an exciton-polariton condensate, here we report the first observation of this mode purely driven by intrinsic quantum fluctuations without such perturbations. By using an ultrahigh quality microcavity and a Raman spectrometer to maximally reject photoluminescence from the condensate, we observe weak but distinct photoluminescence at energies below the condensate emission. We identify this as the so-called ghost branches of the amplitude mode arising from quantum depletion of the condensate into this mode. These energies, as well as the overall structure of the photoluminescence spectra, are in good agreement with our theoretical analysis.**

I. INTRODUCTION

Spontaneous breaking of a continuous symmetry occurs in various branches of modern physics, such as cosmology [1,2], particle physics [3–5], and a variety of condensed matter systems [6–8]. In this broken phase, in addition to the phase (Goldstone) mode [9,10], a collective amplitude (Higgs) excitation [11,12] emerges ubiquitously in various condensates. Such a Higgs amplitude mode has been observed in many condensed matter systems: superconductors [13,14], charge density waves [15], antiferromagnets [16], p-wave superfluids of $^3$He [17], ultracold Fermi superfluid in the Bardeen–Cooper–Schrieffer-Bose Einstein Condensate (BCS-BEC) crossover region [18], bosons loaded in an optical lattice [19], and a supersolid realized in two crossed optical cavities [20]. In all the above observations, it was essential to drive the condensate out of equilibrium to excite and detect this mode.

However, these collective modes are intrinsically driven by quantum fluctuations *without* such external perturbations, even in the ground state where no thermal fluctuations exist. Being in a state of definite phase, interactions in the condensate enable processes that do not conserve the number of condensed particles. As a result, quantum fluctuations coherently expel the particles out of the condensate. This phenomenon, known as quantum depletion, was formulated by Bogoliubov [21] and was recently confirmed in an atomic BEC [22] and an exciton-polariton BEC experiment [23].

The expelled particles occupy the collective modes of the condensate. The spectral signature of this fascinating property is a set of "ghost branches" ($\mathcal{GB}$), the time-reversed partners of the normal collective modes (normal branch or $\mathcal{NB}$) which appear at energies below the ground state. Since the quantum mechanical Bose statistics are crucial in occupying the $\mathcal{GB}$, the observation of these branches provides an unambiguous signature of quantum depletion. Although the ghost branch photoluminescence of the Goldstone mode has been reported [23–25], the detection of the ghost amplitude mode has thus far been experimentally elusive, either due to the property that the amplitude mode is decoupled from density or phase fluctuations in the linear regime in conventional superconductors, or the spectral weight is simply too small to detect.

Microcavity exciton-polariton condensates [26,27] provide a promising testing ground for such quantum many-body phenomena. Exciton-polaritons—half-light, half-matter quasiparticles generated by strong light-matter coupling—inherit a low effective mass and strong interactions from their photonic and excitonic components, respectively, leading to stable BEC at high temperatures. Due to the two-component nature, this system possess an amplitude mode which is an out-of-phase oscillation between the photon and exciton components of the macroscopic wave function [28]. This contrasts with a conventional one-component BEC where no such modes exist.

The polariton system's greatest advantage for the observation of the ghost branch lies in the direct coupling of the photonic component to free photons outside the cavity that can be directly imaged. Whereas atomic BEC experiments rely on time-of-flight or more sophisticated probe techniques to infer the state of the condensate, here the photoluminescence (PL) of the system directly measures the occupation of its photonic component.

## II. HIGH-DYNAMIC RANGE OBSERVATION OF THE GHOST BRANCH AMPLITUDE MODE

Here we report the first direct observation of the quantum-fluctuation-induced amplitude mode, where we observe weak but distinct emission below the condensate emission energy. The energy gap agrees well with the expected amplitude ghost mode energy [28], and the overall structure of the PL spectra is in good agreement with theoretical expectations that we calculate here. Typically, the ghost branch PL is difficult to detect due to the driven-dissipative kinetics[28] and the strong condensate emission that masks the weak spectral weight from the ghost branch. This observation was enabled by ultrahigh-quality microcavities that have increased the cavity lifetime [29] of polaritons that allowed for better thermalized condensates [30], as well as the use of a Raman spectrometer to maximally reject the condensate emission and achieve a dynamic range of 10 orders of magnitude.

Following the methods used by Sun *et al.* [30,31] we form steady-state polariton condensates inside optically generated ring traps. The spectral distribution of the condensate is directly observable in the emitted PL from the microcavity. Unlike previous observations of polariton condensate ghost branches of the Goldstone mode [24,25,32], our condensates are quasi-CW (steady state) and trapped. Excitation densities are on the order of 4 times the condensation threshold, and lower polariton (LP) blueshifts are on the order of 0.5 meV in comparison to a Rabi coupling of ~7 meV. We do not rely on four-wave-mixing to probe the polariton spectrum [32], but instead directly observe PL emission from the microcavity in a stable, steady state with no external perturbing field. Therefore, this experiment probes the intrinsic quantum fluctuations of a condensate, rather than modes that are only observable after driving the system or via indirect probes.

The above-threshold PL is collected from the inside of the pump ring. PL is integrated over the NA of the objective (corresponding to wavenumber region $|k| < k_c = 2.3$ μm$^{-1}$) and a ~0.5 mm x 0.1mm spatial-image plane slit, then spectrally dispersed using a Jobin Yvon T64000 triple Raman spectrometer to achieve maximum rejection of the laser-like condensate emission.

Through a careful use of spectrometer, exposure, and filter settings, we stitched spectra together to achieve a precise dynamic range of over 10 orders of magnitude. Such spectra require long integration times and stability of the system on the order of hours. The overlapping of spectra in Figure 1 is an artifact of the stitching process, and the variation of these data gives an indication of the system stability and precision of the stitching process.

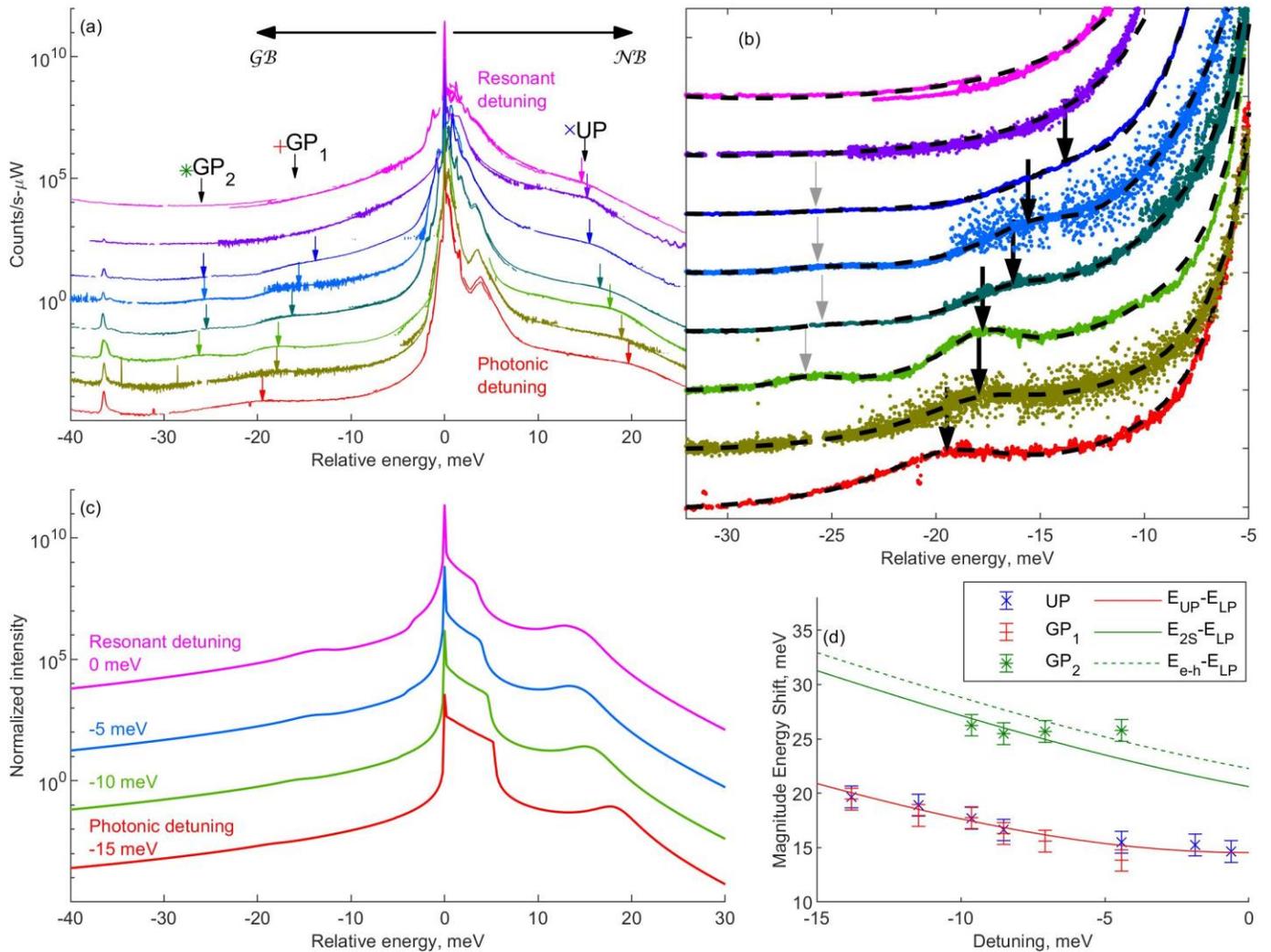

Figure 1: Real and Ghost-branch emission from polariton condensates. Throughout this paper, experimental PL spectra are consistently color-coded across figures and frames. (a) high-dynamic-range photoluminescence spectrally resolved relative to the condensate energy. Energies are defined relative to the condensate energy. (b) Linear-scale zoom-in to the negative energy peaks $GP_1$ and $GP_2$ from frame (a). Data are normalized within the window and offset. Black lines are multi-Lorentzian fits used to identify $GP_1$ and $GP_2$. (c) calculations of the quantum-fluctuation driven ghost branch populations at finite temperature bath (T=3 meV = 30 K). These mirrored peaks at ± 15-20 meV correspond to $GP_1$ and UP in frame (b). (d) magnitude of energy shift of observed peaks in (a) vs detuning. Error bars discussed in the Appendix. The $GP_1$ and UP energies are consistent. Red line: theoretical amplitude mode energy, in agreement with the data. Green lines: $GP_2$ falls at the energy scale for a jump from condensate polariton to cavity-coupled excited excitons (2S state solid green) or unbound e-h plasma (green dashed curve).

## III. QUANTUM FLUCTUATIONS INTO THE AMPLITUDE MODE

Figure 1(a) shows the photoluminescence spectra at different detunings, as a function of the relative energy from the condensate emission energy. All experimental photoluminescence spectra in this paper use the same color-coded spectra for easy comparison. Here, the energy scale is defined relative to the condensate emission at each detuning. Faint but detectable signals on the order of 15-30 meV above and below the condensate signal (plotted respectively as positive and negative energy) are observed. These peaks are assigned according to the discussion in Appendix A. Each spectrum (offset in Figure 1(a) for clarity) is from a different exciton-photon detuning. The sharp peak at -36 meV corresponds to the GaAs phonon replica of the condensate. We observe a positive energy shoulder at the energy of the upper polariton (UP) at 15 to 20 meV. Critically, a negative energy peak ($GP_1$) is observed at the same energy shift below the condensate as the UP is above it. By comparing to the signal strength of the phonon replica, in Appendix G we confirm that the intensity of this $GP_1$ is much stronger than expected for incoherent Raman-like processes.

These features can be interpreted as the appearance of the normal and the ghost amplitude mode in the broken phase. A minimal model to qualitatively explain such behavior is given by the Gross-Pitaevskii (GP) equations [9,33,34] extended to two components to explicitly treat cavity photons and excitons [28,35,36],

$$i\hbar \partial_t \begin{pmatrix} \Psi_c(r,t) \\ \Psi_x(r,t) \end{pmatrix} = M(\nabla) \begin{pmatrix} \Psi_c(r,t) \\ \Psi_x(r,t) \end{pmatrix} = \begin{pmatrix} \hbar\omega_c - \dfrac{\hbar^2 \nabla^2}{2m_c} & g_R \\ g_R & \hbar\omega_X - \dfrac{\hbar^2 \nabla^2}{2m_x} + U_x|\Psi_x(r,t)|^2 \end{pmatrix} \begin{pmatrix} \Psi_c(r,t) \\ \Psi_x(r,t) \end{pmatrix}. \quad (1)$$

Here, $\Psi_{c(x)}(r,t), \hbar\omega_{c(x)}, m_{c(x)}$ are the macroscopic wave function, energy, and effective mass of the cavity photons (excitons), respectively, and $g_R$ is the Rabi splitting. The exciton-exciton interaction $U_x > 0$ gives rise to the blue shift of the exciton level. In equilibrium, the fluctuation-dissipation theorem gives the photoluminescence $[PL]_k(\omega) \propto n_B(\omega) S_k(\omega)$, where $n_B(\omega) = (e^{\hbar\omega/T} - 1)^{-1}$ is the Bose distribution at temperature $T$. The spectrum $S_k(\omega)$ is given by the two-time correlation function of the photonic component, which can be computed by expanding the GP equation in terms of fluctuations $\overrightarrow{\delta\Psi}(k,\omega) = (\delta\Psi_c(k,\omega), \delta\Psi_x(k,\omega), \delta\Psi_c^*(k,\omega), \delta\Psi_x^*(k,\omega))^T$ of the fields $\Psi_{c(x)}$ away from their steady-state values. The fluctuations obey $\omega \overrightarrow{\delta\Psi}(k,\omega) = L_k \overrightarrow{\delta\Psi}(k,\omega)$, where

$$L_k = \begin{pmatrix} M_k + G + E_{LP} & G \\ -G & -M_k - G - E_{LP} \end{pmatrix},$$

$E_{LP}$ is the lower polariton energy, and $G = \text{diag}(0, U_x|\Psi_x^0|^2)$. The spectrum is in the photonic component: $S(k,\omega) = [-\text{Im}[(\omega + i\delta - L_k)^{-1}]]_{11}$ (its explicit form is provided in Eq. (2) In Appendix F).

In the normal state where there is no macroscopic occupation of the ground state such that the nonlinear term is absent ($|\Psi_x(r,t)|^2 = 0$), $L_k$ factorizes into two redundant copies of $M_k$, and $S_k(\omega)$ consists of the familiar upper and lower polariton branches, as plotted in the negative momentum side of Figs. 2(a), (b) (See also Eq. (2) in Appendix F). In the presence of a condensate, however, fluctuations at $k$ and $-k$ are coupled. The lower mode becomes a sound-like phase mode dispersing from the condensate energy while the upper amplitude mode remains gapped and massive, and more dramatically the spectrum acquires ghost partners of these modes mirrored below the condensate emission energy, as shown in the positive momentum side of Figs. 2(a) and (b). These effects are interaction-driven, disappearing if $U_x = 0$, and we emphasize that the ghost branches are physical: despite dispersing below the chemical potential, they are real excitations around the Bogoliubov vacuum of the system, and occupying them raises the total energy.

The asymmetry in the PL between energies above and below the condensate can be understood from the behavior of $n_B(\omega)$. At positive energies the function decays exponentially to zero and the modes are thermally populated up to $\hbar\omega \sim k_B T$ (where $k_B$ is the Boltzmann Constant). At negative energies, the function saturates to a constant so that beyond $\sim k_B T$ away from the condensate energy, the $[PL]_k(\omega) \sim |S_k(\omega)|$ faithfully reflects the spectral weight, making the ghost branches visible.

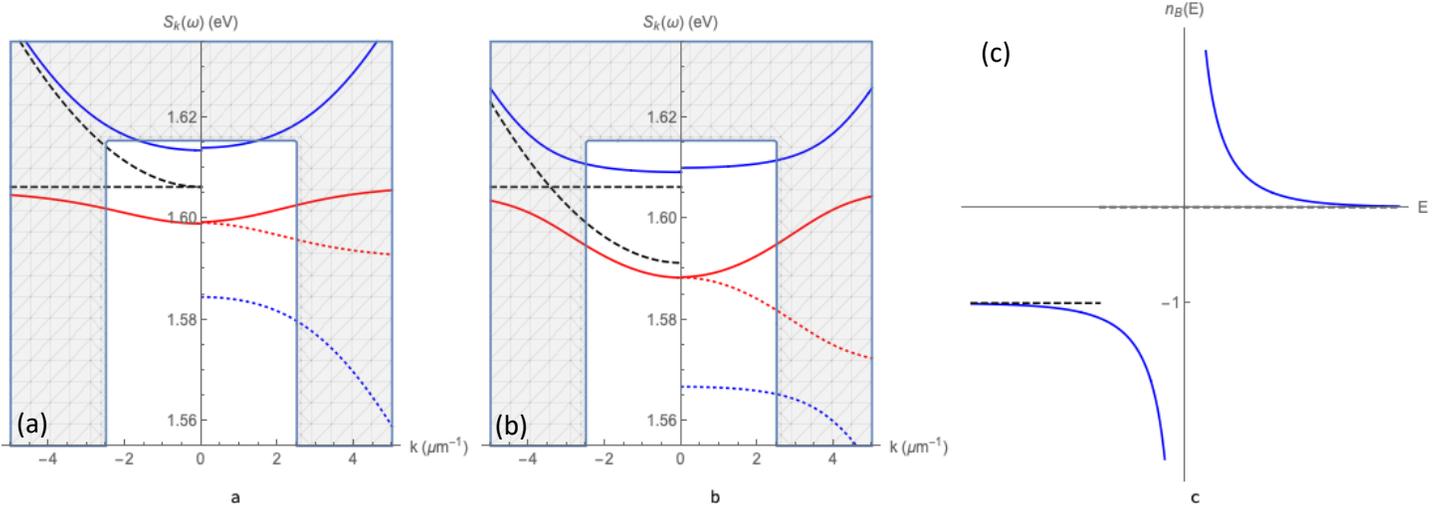

Figure 2: Allowed polariton modes at low density and broken-symmetry regimes. Frames (a) and (b) exciton-polariton system energy branches for resonant detuning and -15 meV photonic detuning, respectively. Normal phase (condensed phase) plotted in negative (positive) wavenumber. Dotted black curves: dispersion of exciton (flat) and photon modes (parabolic). Solid red/blue curves: lower/upper polariton branches. Dashed red/blue curves: ghost branch from the condensed phase. Hashed region: energy/momentum optically excluded in this experiment. Frame (c) Bose distribution function $n_B(\omega)$ showing that the occupation saturates at a constant value at negative energy.

For comparison with the experiment, we have performed a more sophisticated calculation of the photoluminescence spectrum using the theoretical framework based on the Keldysh formalism developed by two of us [37] for a microscopic model of a driven-dissipative electron-hole-photon system, which takes into account driven-dissipative kinetics, thermalization, pair-breaking effects, and finite linewidth (see the Appendix F for the parameters used). As shown in Fig. 1(c), the calculations capture the overall structure of the experimental data. The UP and $GP_1$ energies match the calculated amplitude mode energy at 15-20 meV, with increasing energy for more photonic detunings.

We note that the theory predicts maximum intensity of the $GP_1$ peak at resonance, while our data fails to resolve the peak under these conditions. The data exhibit an increasing broad background near resonance (see Appendix D) that overwhelms the faintly visible $GP_1$. In Appendix F, the theoretical model shows the visibility of this mode as highly dependent on the simplified thermalization parameter. Since this parameter does not catch all the detuning-dependent physics of scattering with the bath, the discrepancy here is not too surprising.

The data in Figure 1(a) also exhibit a faint second negative-energy peak ($GP_2$) between 25 and 30 meV, although we are unable to resolve a positive-energy partner over the exponential roll-off from the UP shoulder. Phenomenologically, this can be understood as the LP condensate coupling to higher energy states. The heavy-hole excitons in our ~7 nm GaAs/AlAs quantum wells should have a roughly 10-14 meV binding energy [38,39], and therefore a closely-spaced ladder of *s*-like exciton states beginning roughly 10 meV above the 1*s* and merging into the electron-hole continuum. The Rabi coupling to these higher-lying states is rapidly suppressed due to the exciton wave function while the density of states increases, producing a broad feature that merges into the continuum, in a manner reminiscent of exciton absorption described by the Elliott formula [40]. The spectral weight in higher-lying states drops as the cavity becomes more detuned from them, resulting in a feature that disperses between the 2*s* state and the continuum as shown in Figure 1(b). The theory of Figure 1(c) does not treat the long-range Coulomb interaction (replacing it with a contact interaction between fermions), and therefore is not expected to capture additional exciton bound states or Coulomb correlations in the electron-hole plasma that lead to this peak in the emission before the onset of the continuum.

The concept of the amplitude mode is developed in terms of the single, complex-scalar order parameter of the polariton condensate system. As stated previously, this has been predicted to manifest as coupling directly to the UP, consistent with the energy scale of our $GP_1$ mode. Just as the Standard Model [3] predicts a single, elementary Higgs boson, this derivation predicts one amplitude mode; however, if the physics of the underlying system is sufficiently rich, then the amplitude mode theory can be extended (c.f. the composite Higgs mode theories, both before [41,42] and since [43,44] the 2012 observation at LHC). $GP_1$ and $GP_2$ both give rise to fluctuations out of the condensate into gapped states—each representing a discrete frequency of oscillation of the particle number. As such, both modes can be viewed as "generalized amplitude modes" relevant to other condensate fields, including theories beyond the Standard Model.

## IV. LOW-ENERGY EXCITATIONS OCCUPIED BY THERMAL DEPLETION

Finally, we examine a family of distinct features we observe in a much narrower energy range, within 3 meV of the condensate energy, plotted in Fig. 3. Much like the collective modes discussed above, these peaks appear as mirrored pairs around the condensate energy with differences in intensity that can be understood from the Bose distribution function at positive and negative energies. Rather than originating from the homogeneous polariton condensate, we interpret these features as arising from the discrete levels of the trapped polariton condensate as well as the thermalization dynamics of hot, untrapped polaritons. These features are particular to our experimental configuration and are not captured by the general theory of Fig. 1(c).

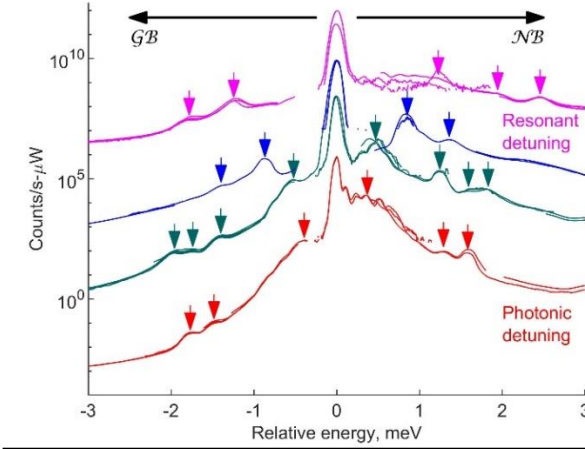

*Figure 3: Spectral features close to the condensate energy. This is subset of the data in Fig1(a) showing mirrored spectral features within the energy range of the LP and corresponding ghost branch. Colors for spectra are maintained between Fig 1 and here. Aside from the instability shifted peaks in the resonant detuning data, the other spectra exhibit well-mirrored peaks. The energy spacing of these levels are consistent with polaritons in our ring traps.*

To understand the origin of these peaks we consider our trapped condensate: to achieve sufficient trapped densities, the optical pump diameter and focus was adjusted for each measurement position. (Details in Appendices A and B.) For trapped condensates on the order of 5-10 μm diameter and a polariton on the order of $3\times10^{-5}$ times the bare electron mass, the observed energy spacing of ≤ 1 meV in Figure 1(d) is consistent with quantum confinement in an approximately square well or harmonic potential. For a comprehensive phase map of excited states present in polariton ring condensates, see Sun et.al. [31] Here we work well into the "Plateau" and "Single" phase regions, but as that reference points out, the boundaries are continuous. Thus, the faintly occupied excited states in those regions of phase space are being detected due to high dynamic range collection and effective rejection of the spectrally-near condensate emission.

Much like previous theory [33,45] and experiment [24,25,32] which focused on the continuous Bogoliubov dispersions that exist in an infinite polariton condensate, the current theory of Figure 1(c) does not account for confinement effects and so cannot be expected to generate the observed features in the spectrum. Qualitatively, however, their interpretation is similar to the continuous modes. The trap gives rise to additional discrete features in the density of states, such as those into which condensation has been observed [31,46]. These are populated to different extents by thermal occupation and quantum depletion, and so produce mirrored peaks of different magnitude on opposite sides of the condensate emission. A more detailed theoretical and experimental study of the fluctuations of the condensate into these discrete states would be interesting in its own right, but this is beyond the scope of the current work.

## V. CONCLUSIONS

We have observed PL from a long-lifetime microcavity polariton condensate that is consistent with quantum fluctuations from the condensate into its collective amplitude modes. A pair of positive-negative peaks (UP and $GP_1$) are seen to agree with the predictions of our random-phase-approximation based theory that includes driven-dissipative kinetics and thermalization effects; this amplitude mode approximately tracks the UP-LP energy gap as we vary the exciton-photon detuning which is consistent with long-standing predictions of the amplitude mode in polariton condensates. Moreover, this state is observed from direct photoluminescence emission from an unperturbed, steady state condensate.

At slightly more negative energy shifts, we observe a second ghost branch peak, $GP_2$. The energy scale of this state suggests it may possibly correlate with the manifold of excited hydrogenic exciton levels, although such physics is not included in the theory that we use to interpret the lower energy peak.

Additionally, we sometimes observe a ladder of states within 2-3 meV of the condensate. The energy spacing of the states, and their occasional appearance, is well explained as the excited states of an optically-confined ~5 µm polariton condensate. Whenever these states are clearly observed at positive energy, there are corresponding negative energy peaks. While current theory and experiment search for a Bogoliubov continuum within this energy scale of a polariton condensate, these results suggest that gapped amplitude modes may emerge from the trapped polariton spectrum.

While the standard derivation of the Higgs amplitude mode for a simple condensate predicts only a single amplitude mode, a more complicated system with numerous branching interactions must have a family of excitation modes for the condensate. Such modes allow for the fluctuation of the condensate particle number at different frequencies, and each can therefore be viewed as a sort of "generalized amplitude mode" in similar fashion that composite Higgs boson theories are invoked to extend beyond the Standard Model.

## VI. ACKNOWLEDGEMENTS

This work was authored in part by the National Renewable Energy Laboratory, operated by Alliance for Sustainable Energy, LLC, for the U.S. Department of Energy (DOE) under Contract No. DE-AC36-08GO28308. Funding provided by U.S. Department of Energy, Office of Science, Basic Energy Sciences (DE-AC36-08GO28308). Work at Argonne supported by DOE office of science basic energy sciences, materials science and engineering under contract DE-AC02-06CH11357. R. H. was supported by a Grand-in-Aid for JSPS fellows (Grant No. 17J01238). The work of sample fabrication at Princeton was funded by the Gordon and Betty Moore Foundation (GBMF-4420) and by the National Science Foundation MRSEC program through the Princeton Center for Complex Materials (DMR-0819860).

## APPENDIX A: SAMPLES AND EXPERIMENTAL METHODS

GaAs-based microcavities used here are the same high Q-factor design as that studied in [29]. Such long cavity lifetimes increase accumulation of polariton density for a given pump fluence and make condensation more accessible. Moreover, the increased ratio of lifetime to relaxation time results in better equilibrated condensates [30].

These microcavity structures include a tapered thickness that results in a gradual variation of the exciton-photon detuning vs sample position. This enables probing different detuning (and therefore polariton character) by translating the sample.

Using a spatial light modulator (SLM) we imprint an axicon phase profile onto a stabilized pump laser tuned to about 100 meV higher energy than the lower polariton state. This ensures that any coherence and polarization of the exciting laser is lost as carriers cool into the polariton states. The ring diameter and focus can be adjusted on-the-fly using the SLM and intermediate optics, and the carriers generated in the ring create a barrier in the energy landscape of the lower polariton, causing polaritons to accumulate inside the ring. To mitigate laser heating, we chop the CW laser at 360 Hz with a 13.7% duty cycle. We pump at normal incidence and collect the photoluminescence through the same 0.28 NA objective.

Spectra are collected on a Jobin Yvon T64000 triple Raman spectrometer to achieve required spectral rejection. High dynamic range spectra are assembled from combining data from a wide range of exposure and filter settings as well as carefully selected grating settings.

## APPENDIX B: RING SIZE AND RESOLUTION OF TRAPPED STATES

The energy spacing resolved in Fig. 3 is consistent with a polariton condensate trapped in a ring diameter on the order of 10 µm, consistent with the conditions of this study. Too small of a ring diameter (when ring diameter ~ pump focus resolution or exciton diffusion) washes out the trapping potential and the polaritons become untrapped. At the other limit, too large of a ring diameter will reduce the local excitation density along the ring perimeter, reducing the potential energy barrier and the polaritons will again become untrapped (starting with the excited states). Additionally, local defects can perturb the condensate. In particular, our structures exhibit accumulated strain in the AlAs/AlGaAs DBRs exhibit a network of defect planes (such as seen in Fig. 4) that appear to act as line barriers to the polaritons as seen in other strained systems [47]. This network of defects limited the maximum condensate size in some regions of the sample in order to exclude defects from the ring.

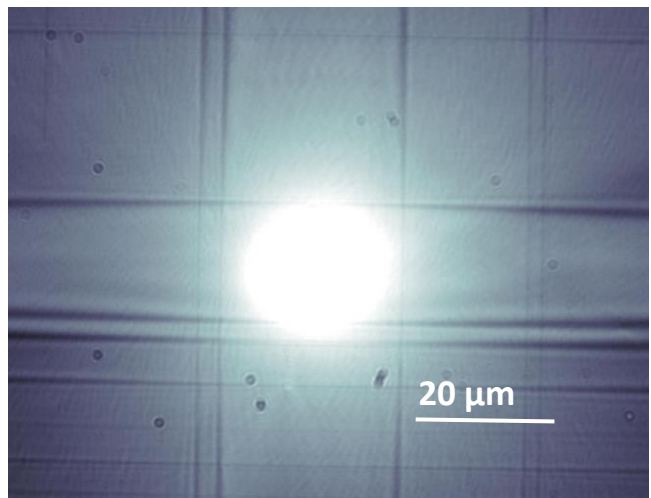

*Figure 4: propagation of polaritons away from the pump region shows line defects. The white region includes the optical pump and is highly saturated. If a pump ring crosses a significant defect line, the resulting condensate can be disturbed or segmented.*

We only observe these peaks in about half of the spectra taken. Their appearance does not correlate with detuning (peaks were resolved over the entire range of detunings studied). Thus, we attribute their appearance as hypersensitive to the trap diameter, barrier height, and local defects. If the trap is deep enough to support excited bound states, and the interlevel splitting is large enough to resolve, then we should expect to observe a ladder of thermally or kinetically-occupied excited states in the condensate. As ring diameter was optimized only to achieve condensation at each detuning and avoid defect lines, it should not be surprising that we only sometimes fall within the "goldilocks zone" of parameters that resolve these trapped states.

## APPENDIX C: QUANTITATIVE ASSIGNMENT OF UP, $GP_1$, $GP_2$, AND RELATED ERROR BARS

To extract the negative peaks, we do a least-squares fit to the logarithmic data in the energy range of -40meV <= $\Delta E$ <=-5meV after excluding outliers due to CCD defects or the phonon replica. It is critical to heavily weight the data or calculate the error on a logarithmic scale to be sensitive to features over 10 orders of magnitude. The background is modeled with three Lorentzians: one to account for the condensate emission, one to model the turn up at negative energies (due to substrate emission, see Appendix D), and a third at $\Delta E$>-5 meV to allow for the manifold of excited states seen around the condensate. The modes of interest, $GP_1$ and $GP_2$ are modeled as Lorentzian lineshapes, only including $GP_2$ when the data merits it. Alternate fitting models (not shown) were tested in addition to the above empirical prescription in order to test the validity of these fits. The results of the fits are qualitatively in agreement, and the variation between the fit results was used to set the error bars in Fig 1d. Qualitatively, this error accounts for the uncertainty in the background model and the range of data that the user includes in the fitting.

We identify a positive energy shoulder (UP) in the range of 15-20 meV above the condensate. At higher energy shift, the PL counts decrease with a much steeper exponential falloff, and the energy of this shoulder is consistent with the energy of the upper polariton at each detuning measured.

## APPENDIX D: VISIBILITY OF $GP_1$ AND $GP_2$ AND SUBSTRATE LUMINESCENCE

The theory (Fig. 1(c)) clearly indicates that the amplitude mode ($GP_1$) should exhibit a larger intensity near resonance and decrease at photonic detunings. However, the data of Fig. 1(a,b) only resolve this mode at moderate photonic detunings. We observed an unexplained broad background that increased in amplitude close to resonance, seen clearly when we plot absolute counts/s in Fig. 5. Here the background at resonance is 1-2 orders of magnitude higher than at photonic detunings, which obscures $GP_1$ and $GP_2$ just as it obscures the phonon replica at -36meV. The origin of this background isn't well understood, but it could be related to the intrinsic linewidth of the condensate (i.e. due to disorder in the exciton), spectrometer broadening (e.g. spectrometer slit widths), or due to an encroaching pump barrier (pump ring diameter had to be reduced to encourage condensation of less-mobile excitonic polaritons). Also see Appendix F for an analysis of the lineshapes of the theory results and how these are impacted by the thermalization parameter.

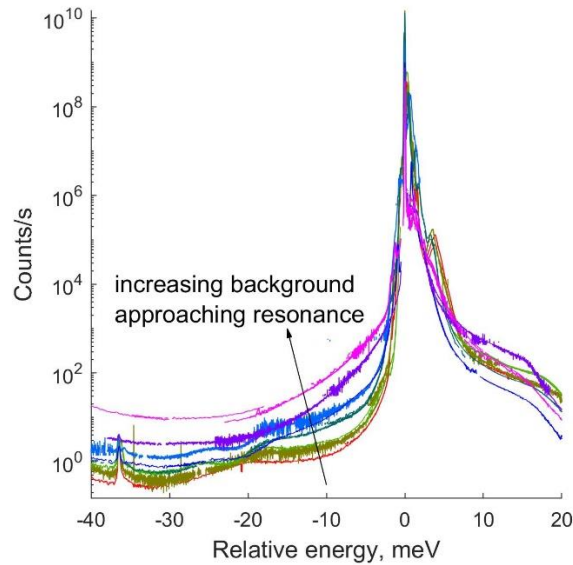

*Figure 5: Absolute counts/s of PL spectra for all detunings. The most photonic detuning (red) exhibits the lowest background in the region of $GP_1$ and $GP_2$ while the background increases significantly near resonance.*

When fitting for $GP_1$ and $GP_2$ (see Appendix C), we must account for luminescence from excited carriers in the substrate that is filtered by the long-wavelength transmission of the distributed Bragg reflector (DBR). This is seen in Fig. 6. Frame (a) shows a calculated transmission curve for the DBR in our device. Frame (b) shows a representative spectrum extended out to more than 100 meV red-detuned from the condensate. The polariton states exist within the highly reflective stopband of the DBR, while the hot substrate recombination can leak through the cavity outside that stopband. Here we only claim a qualitative agreement because the transfer matrix calculation only uses an approximate index model for the materials.

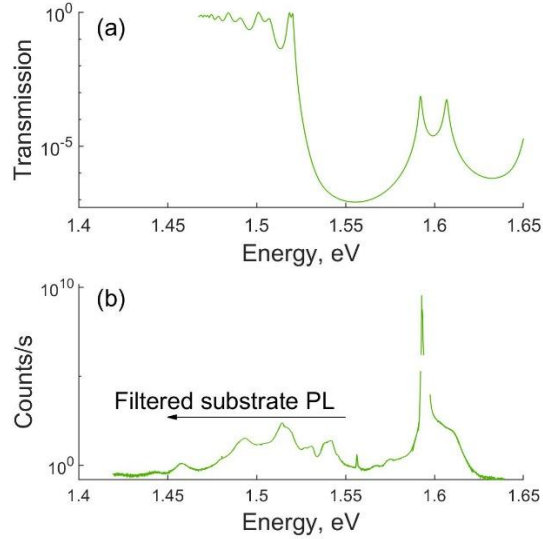

Figure 6: Long-wavelength photoluminescence from the substrate leaks through the DBR of the cavity. Frame (a): calculated transmission of the DBR. Frame (b): Representative broadband PL spectra for moderately photonic data. Emission begins to leak through the stopband of the DBR below about 1.55 eV. This is in qualitative agreement with the cavity transmission.

## APPENDIX E: TWO-COMPONENT GROSS-PITAEVSKII EQUATION

The steady state of the two-component Gross-Pitaevskii equation (1) is given by the ansatz, $\Psi_{c(x)}(r,t) = \Psi_{c(x)}^0 e^{-\frac{iEt}{\hbar}}$, where the condensate emission energy $E$ should equal the (blue shifted) lower or the upper polariton eigenenergy,

$$E_{UP/LP} = \frac{1}{2}\left[\hbar\omega_c + \hbar\omega_x + U_x|\Psi_x^0|^2 \pm \sqrt{(\hbar\omega_c - \hbar\omega_x - U_x|\Psi_x^0|^2)^2 + 4g_R^2}\right].$$

From physical perspective, $E = E_{LP}$ [37]. Linearizing the GP equation (1) around the equilibrium state $\Psi_{c/x}(r,t) = e^{-\frac{iEt}{\hbar}}[\Psi_{c/x}^0 + \delta\Psi_{c/x}(k,\omega)e^{ik\cdot r - i\omega t} + \delta\Psi_{c/x}^*(-k,-\omega)e^{-ik\cdot r + i\omega t}]$, we obtain $\omega\delta\vec{\Psi}(k,\omega) = L_k \delta\vec{\Psi}(k,\omega)$, where

$$L_k = \begin{pmatrix} K_k & G \\ -G & -K_k \end{pmatrix} = \frac{1}{\hbar}\begin{pmatrix} \hbar\omega_c + \frac{\hbar^2 k^2}{2m_c} - E_{LP} & g_R & 0 & 0 \\ g_R & \hbar\omega_x + \frac{\hbar^2 k^2}{2m_x} + 2U_x|\Psi_x^0|^2 - E_{LP} & 0 & U_x|\Psi_x^0|^2 \\ 0 & 0 & -\hbar\omega_c - \frac{\hbar^2 k^2}{2m_c} + E_{LP} & -g_R \\ 0 & -U_x|\Psi_x^0|^2 & -g_R & -\hbar\omega_x - \frac{\hbar^2 k^2}{2m_x} - 2U_x|\Psi_x^0|^2 + E_{LP} \end{pmatrix},$$

and $\delta\vec{\Psi}(k,\omega) = (\delta\Psi_c(k,\omega), \delta\Psi_x(k,\omega), \delta\Psi_c^*(k,\omega), \delta\Psi_x^*(k,\omega))^T$. Here, we have assumed $\Psi_x^0 > 0$ to be real without loss of generality.

The first two lines involve the propagating fluctuations $\delta\Psi_{c/x}(k,\omega)$, while the last two lines involve the *counter-propagating* fluctuations $\delta\Psi_{c/x}^*(k,\omega)$. In experiments, we have access to the propagating fluctuations in the photonic component, where the spectral function is given by

$$S(\boldsymbol{k}, \omega) = \left[-\text{Im}[(\omega + i\delta - L_{\boldsymbol{k}})^{-1}]\right]_{11}$$

$$= -\text{Im} \frac{K_{\boldsymbol{k}22} \det(\omega + i\delta + K_{\boldsymbol{k}}) - K_{\boldsymbol{k}11}(U_x |\Psi_x^0|^2)^2}{\det(\omega + i\delta - K_{\boldsymbol{k}}) \det(\omega + i\delta + K_{\boldsymbol{k}}) + \left(U_x |\Psi_x^0|^2\right)^2 \left((\omega + i\delta)^2 - K_{\boldsymbol{k}11}^2\right)}. \quad (2)$$

In the normal state $|\Psi_c^0|^2 = |\Psi_x^0|^2 = 0$, $L_{\boldsymbol{k}}$ consists of two redundant matrices on the block diagonal; as can be seen explicitly in Eq. (S1), this redundancy cancels in the spectrum and only the positive-frequency set of propagating modes is visible. In this case, the spectrum $S(\boldsymbol{k}, \omega)$ simply exhibits a delta function peak at the normal mode,

$$S(\boldsymbol{k}, \omega) = |C_{\boldsymbol{k}}|^2 \delta(\omega - \hbar\omega_{\boldsymbol{k}}^-) + |X_{\boldsymbol{k}}|^2 \delta(\omega - \hbar\omega_{\boldsymbol{k}}^+) \quad (3)$$

where

$$\hbar\omega_{\boldsymbol{k}}^{\pm} = \frac{1}{2}\left[\hbar\omega_{\boldsymbol{k}}^c + \hbar\omega_{\boldsymbol{k}}^x \pm \sqrt{(\hbar\omega_{\boldsymbol{k}}^c - \hbar\omega_{\boldsymbol{k}}^x)^2 + 4g_R^2}\right].$$

The Hopfield coefficients are

$$|C_{\boldsymbol{k}}|^2 = 1 - |X_{\boldsymbol{k}}|^2 = \left[1 - \frac{\delta_{\boldsymbol{k}}}{\sqrt{\delta_{\boldsymbol{k}}^2 + 4g_R^2}}\right]/2$$

with $\hbar\omega_{\boldsymbol{k}}^{c(x)} = \hbar\omega_{c(x)} + \frac{\hbar^2 k^2}{2m_{c(x)}} - E$ and $\delta_{\boldsymbol{k}} = \hbar\omega_{\boldsymbol{k}}^c - \hbar\omega_{\boldsymbol{k}}^x$.

When the exciton-exciton interaction is turned on $U_x > 0$, in contrast, the propagating and the counter-propagating fluctuations couple, making the eigenmode a mixture of the propagating and counter-propagating fluctuations. This allows the ghost mode in the spectrum $S(\boldsymbol{k}, \omega)$, where the ghost mode eigenenergies are given by,

$$\hbar\omega_{\boldsymbol{k},G}^{\pm} = -\frac{1}{2}\left[A_{\boldsymbol{k}} \pm \sqrt{A_{\boldsymbol{k}}^2 - 4B_{\boldsymbol{k}}}\right],$$

in addition to the normal branch $\hbar\omega_{\boldsymbol{k},N}^{\pm} = -\hbar\omega_{\boldsymbol{k},G}^{\pm}$, where

$$A_{\boldsymbol{k}} = (\hbar\omega_{\boldsymbol{k}}^c)^2 + (E_{\boldsymbol{k}}^x)^2 - 2g_R^2, B_{\boldsymbol{k}} = (\hbar\omega_{\boldsymbol{k}}^c)^2 (E_{\boldsymbol{k}}^x)^2 - 2g_R^2 \hbar\omega_{\boldsymbol{k}}^c (\hbar\omega_{\boldsymbol{k}}^x + 2U_x |\Psi_x^0|^2) + 4g_R^2,$$

and $(E_{\boldsymbol{k}}^x)^2 = (\hbar\omega_{\boldsymbol{k}}^x + 2U_x)^2 - U_x |\Psi_x^0|^2$, plotted in Fig. 2. As originally pointed out in Ref. [28], the mode with the eigenenergy $\hbar\omega_{\boldsymbol{k},G}^+$, is associated with the amplitude fluctuations, which is the ghost amplitude mode GP$_1$, our central scope of this paper.

## APPENDIX F: MICROSCOPIC ANALYSIS

Our microscopic theoretical analysis performed in the main text follows the method developed by two of us in Ref. [37]. In this section, we provide a brief description of the model and the method of our microscopic analysis, along with the parameters used in Fig. 1(c) in the main text. We emphasize that the parameters used in our microscopic model are directly extracted from the measurements, except for one phenomenological parameter that controls the linewidth of the spectrum. For more details, such as the explicit expression of our Hamiltonian $H$ and the computation method, we refer to Ref. [37].

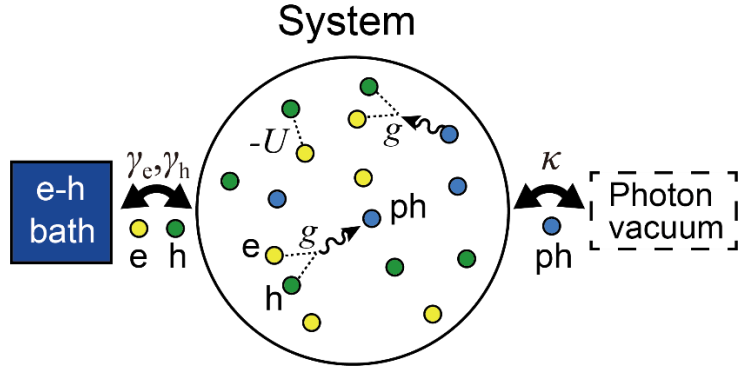

*Figure 7: Model driven-dissipative electron-hole-photon gas, where the system is attached to an electron-hole bath and a photon vacuum. Electrons (holes) are incoherently supplied to the system with the rate $\gamma_{e(h)}$ (where we put $\gamma_e = \gamma_h = \gamma$ for simplicity). In the system, the injected electrons (``e'') and holes (``h'') attractively (e-h) interact with a contact-type interaction with coupling strength $-U$. The electrons and holes pair-annihilate (create) to create (annihilate) cavity-photons (``ph'') via the dipole coupling $g$. The created photons in the cavity leak out to the vacuum with the decay rate $\kappa$.*

Our model is schematically shown in Fig. 7. Here, the system, which consists of electrons, holes, and photons, is attached to a photon vacuum and an electron-hole bath. The coupling between the electron-hole bath and the system enables the electrons and holes to be injected into the system and get thermalized with the rate $\gamma_e$ and $\gamma_h$, respectively (where we set $\gamma_e = \gamma_h = \gamma$ for simplicity). The injected electrons and holes, which have the kinetic energy $\varepsilon_{\boldsymbol{p}} = \frac{\hbar^2 \boldsymbol{p}^2}{2m_{\text{eh}}} + \frac{E_g}{2}$ (where $m_{\text{eh}}$ is the effective mass of the electron and the hole that is assumed to have the same mass for simplicity and $E_g$ is the energy gap of the semiconductor material), pair-create/pair-annihilate photons via the dipole coupling $g$ in the system. The created photons, which have the kinetic energy $\varepsilon_{\boldsymbol{k},ph} = \hbar\omega_{\text{cav}} + \frac{\hbar^2 k^2}{2m_{\text{cav}}}$ (where $m_{\text{cav}}$ is the cavity photon effective mass), leak out to the photon vacuum.

The electron-hole bath and the photon vacuum are assumed to be large enough such that they stay in equilibrium. The bath electron-hole distribution is characterized by the bath chemical potential $\mu_b$ and temperature $T_b$, given by ($k_B$ is the Boltzmann's constant),

$$f_b(\omega) = \frac{1}{e^{[\hbar\omega - (\mu_b + E_g/2)]/(k_B T_b)} + 1}.$$

Here, the bath chemical potential $\mu_b$ controls the pumping rate. The occupation of the photon vacuum is absent, i.e. $f_v(\omega) = 0$. With this setup, the system eventually reaches a nonequilibrium steady state by reaching a balance between the electron-hole pumping and the photon decay.

In the system, the electrons and holes Coulomb interact with each other to form an exciton in the dilute limit at the energy level $\hbar\omega_X = E_g - E_{\text{bind}}^X$, where $E_{\text{bind}}^X$ is the binding energy of the exciton. This enables us to define the detuning parameter $\delta = \hbar\omega_{\text{cav}} - \hbar\omega_X$. We briefly note that, in our analysis, we have assumed, for simplicity, a contact-type interaction with a coupling constant $-U < 0$, instead of the realistic long-range Coulomb type interaction between the electrons and holes. We expect this simplification to have only a little impact on the excitation properties (except for the rise of the secondary ghost peaks (GP$_2$) that we argue to be originated from the higher order Rydberg series), at least in the low density regime, where the detailed properties of the attractive interaction that binds the electron and the hole would not be so important.

To compute the photoluminescence spectrum of the above model, we have performed a generalized random phase approximation combined with the Hartree-Fock-Bogoliubov approximation [37]. In this approach, we first determine the nonequilibrium steady state within the meanfield approximation for a given parameter set, and then calculate the fluctuations around the obtained steady state that directly relates to the photoluminescence spectrum,

$$[PL]_{\boldsymbol{k}}(\omega) = \frac{1}{2\pi}\int_{-\infty}^{\infty} d(t-t')e^{-i\omega(t-t')}\langle a_{\boldsymbol{k}}^{\dagger}(t')a_{\boldsymbol{k}}(t)\rangle,$$

where $a_{\boldsymbol{k}}$ is an annihilation operator of a cavity photon. These are computed to be consistent with the meanfield approximation used for the computation of the steady state, by utilizing the Keldysh diagrammatic techniques. We refer to Ref. [37] for details.

In our measurement, the photoluminescence spectra are collected and integrated over the NA of the objective. This corresponds to integrating $[PL]_{\boldsymbol{k}}(\omega)$ over momenta as,

$$\frac{1}{2\pi}\int_0^{k_c} dk\, k[PL]_{\boldsymbol{k}}(\omega),$$

with a sharp cutoff $k_c = 2.3\ \mu m^{-1}$.

Below, we list the parameters used in our calculation, that are chosen to be as consistent to the experiment in the main text as possible. We have chosen $U = 5.2$ meV/μm^2 to reproduce the exciton binding energy $E_{\text{bind}}^{\text{X}} = 10$ meV, where we have used the parameter $m_{\text{eh}} = 0.068 m_0$ ($m_0$ is the electron mass). Similarly, we have chosen $g = 1.7$ meV/μm$^2$ to reproduce the measured Rabi splitting of $2g_R = 14$ meV in the dilute limit. (See the Appendix A in Ref. [37] for more details.) We set the photon decay rate to $\kappa = 0.03$ meV, which corresponds to the cavity photon lifetime of $\tau = 140$ ps. The photon mass is $m_{\text{cav}} = 3 \times 10^{-5} m_0$ where $m_0$ is the mass of the bare electron. The photon density is fixed to be $n_{ph} = 180\ \mu m^{-2}$, which gives a blue shift to the condensate of order $\sim 0.1 - 1$ meV.

The phenomenological parameter $\gamma$ that roughly corresponds to the thermalization rate, unfortunately, cannot be directly extracted from the experimental data. Here, we provide in Fig. 8 the $\gamma$-dependence of the photoluminescence spectrum. As seen here, the ghost branch can easily be masked by the linewidth (see Appendix D). In Fig. 1(c), we have set $\gamma = 4$ meV to yield a comparable overall structure to the experiment.

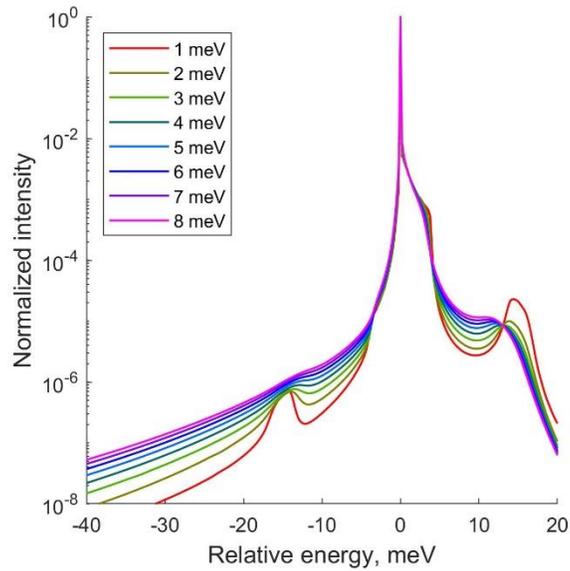

*Figure 8: γ-dependence of the photoluminescence spectrum. Data calculated at resonant detuning. Here γ is related to the thermalization rate, so that higher values result in a better thermalized condensate.*

APPENDIX G: COMPARISON OF GP$_1$ TO INCOHERENT RAMAN SCATTERING

In the main text, we noted that a Raman-like Stokes line appears for the longitudinal optical (LO) phonon at an energy 36 meV below the polariton ground state, with a spectral width matching that of the condensate. The question naturally arises whether the other lines below the polariton ground state could be interpreted similarly. In the case of the phonon Raman-like line, a phonon is emitted during the photon emission process. In principle, the same could occur for electronic excitations, e.g., a second polariton could be kicked up to a higher excited state during the photon emission process. Such a process can occur for incoherent, non-condensed polaritons.

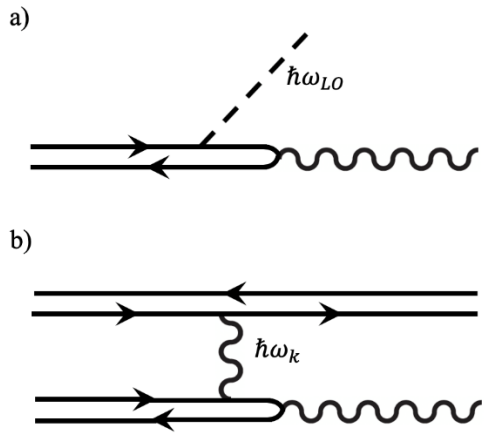

*Figure 9: Interaction diagrams for processes we compare: (a) polariton-optical phonon emission and (b) polariton-polariton scattering.*

To estimate the likelihood of such a process, we can use the optical phonon Stokes line as a benchmark and compare the cross-section for incoherent electronic excitations to this. The scattering diagram for the optical phonon emission process is shown in Fig. 9(a). We assume that the momentum of the polaritons is negligible, i.e. in the ground state or nearby. Following the rules for Rayleigh-Schrödinger diagrams (see, e.g., Ref. [48], chapter 8), we write the rate for this process as

$$\frac{1}{\tau} = \frac{2\pi}{\hbar} \sum_{\vec{k}} |M|^2 \delta(E_{pol}(0) - \Delta E - \hbar ck),$$

where $\Delta E = \hbar \omega_{LO}$, and we sum over the two-dimensional range of in-plane $k$-vectors of the emitted photons, constrained by Snell's law that the in-plane k-component of the internal polariton state must match the in-plane component of the external photon. For the matrix element we use the standard Fröhlich interaction vertex, adjusted for two dimensions, since the polaritons are constrained to move only in the cavity plane. This gives us

$$|M_{Fr}|^2 = \frac{1}{A} \frac{e^2}{\epsilon k} \frac{\hbar \omega_{LO}}{\left(\Delta E + \frac{\hbar^2 k^2}{2m}\right)^2} M_{Phot}^2$$

where $e$ is the electron charge and $\epsilon$ is the effective permittivity, and $M_{Phot}$ is the photon emission vertex.

By comparison, for the process shown in Figure 9(b), we assume a hard-core interaction for the polaritons, in which case the matrix element is

$$|M_{coll}|^2 = \frac{1}{A} \frac{ng^2}{\left(\Delta E + \frac{\hbar^2 k^2}{2m}\right)^2} M_{Phot}^2$$

where $n = N/A$ is the density of polaritons and $g$ is the interaction strength. The value of $g$ ranges in the literature from 0.001 to 40 μeV-μm².

Taking $\Delta E$ and k to the same order of magnitude in both cases, this gives us the ratio

$$\frac{|M_{coll}|^2}{|M_{Fr}|^2} = \frac{ng^2}{(e^2/\epsilon)\hbar \omega_{LO}} (E_{pol}/\hbar c)$$

where have substituted $k \sim E_{pol}/\hbar c$. Using $\hbar \omega_{LO} = 36$ meV, taking the upper bound for the value of g given above, and using the polariton density $n \sim 10^8$ cm$^{-2}$ for a typical polariton density near condensation, we obtain

$$\frac{|M_{coll}|^2}{|M_{Fr}|^2} \sim 10^{-4}$$

In other words, if the extra lines are Raman-like, incoherent scattering processes, they should have spectral weight about 10$^{-4}$ times the spectral weight of the LO phonon line. In the experiments, however, these lines have roughly the same or greater spectral weight as the phonon line, indicating that they cannot be explained this way, and that the condensate properties are crucial for understanding them.

## References

[1]   D. Kazanas, *Dynamics of the universe and spontaneous symmetry breaking,* Astrophys. J. **241**, L59 (1980).

[2]   W. H. Zurek, *Cosmological experiments in condensed matter systems,* Phys. Rep. **276**, 177 (1996).

[3]   P. W. Higgs, *Broken symmetries and the masses of gauge bosons,* Phys. Rev. Lett. **13**, 508 (1964).

[4]   P. Higgs, *Prehistory of the Higgs boson,* Comptes Rendus Phys. **8**, 970 (2007).

[5]   P. W. Anderson, *Superconductivity: Higgs, Anderson and all that,* Nat. Phys. **11**, 93 (2015).

[6]   Y. Nambu, *Quasi-particles and gauge invariance in the theory of superconductivity,* Phys. Rev. **117**, 648 (1960).

[7]   J. Goldstone, A. Salam, and S. Weinberg, *Broken symmetries,* Phys. Rev. **127**, 965 (1962).

[8]   A. Edelman and P. B. Littlewood, *BEC to BCS Crossover from Superconductors to Polaritons,* in *Univers. Themes Bose-Einstein Condens.*, edited by N. P. Proukakis, D. W. Snoke, and P. B. Littlewood (Cambridge University Press, 2017), pp. 1–9.

[9]   M. Wouters and I. Carusotto, *Excitations in a Nonequilibrium Bose-Einstein Condensate of Exciton Polaritons,* Phys. Rev. Lett. **99**, 140402 (2007).

[10]  T. W. Chen, S. C. Cheng, and W. F. Hsieh, *Collective excitations, Nambu-Goldstone modes, and instability of inhomogeneous polariton condensates,* Phys. Rev. B **88**, 184502 (2013).

[11]  P. B. Littlewood and C. M. Varma, *Amplitude collective modes in superconductors and their coupling to charge-density waves,* Phys. Rev. B **26**, 4883 (1982).

[12]  D. Pekker and C. M. Varma, *Amplitude / Higgs Modes in Condensed Matter Physics,* Annu. Rev. Condens. Matter Phys. **6**, 269 (2014).

[13]  R. Sooryakumar and M. V. Klein, *Raman scattering by superconducting-gap excitations and their coupling to charge-density waves,* Phys. Rev. Lett. **45**, 660 (1980).

[14]  R. Matsunaga, Y. I. Hamada, K. Makise, Y. Uzawa, H. Terai, Z. Wang, and R. Shimano, *Higgs amplitude mode in the bcs superconductors Nb1-xTi xN induced by terahertz pulse excitation,* Phys. Rev. Lett. **111**, 057002 (2013).

[15]  J. Demsar, K. Biljakovic, and D. Mihailovic, *Single Particle and Collective Excitations in the One-Dimensional Charge Density Wave Solid K0.3MoO3 Probed in Real Time by Femtosecond Spectroscopy,* Phys. Rev. Lett. **83**, 800 (1999).

[16]  C. Rüegg, B. Normand, M. Matsumoto, A. Furrer, D. F. McMorrow, K. W. Krämer, H. U. Güdel, S. N. Gvasaliya, H. Mutka, and M. Boehm, *Quantum magnets under pressure: Controlling elementary excitations in TlCuCl3,* Phys. Rev. Lett. **100**, 205701 (2008).

[17]  V. V Zavjalov, S. Autti, V. B. Eltsov, P. J. Heikkinen, and G. E. Volovik, *Light Higgs channel of the resonant decay of magnon condensate in superfluid (3)He-B,* Nat. Commun. **7**, 10294 (2016).

[18]  A. Behrle, T. Harrison, J. Kombe, K. Gao, M. Link, J. S. Bernier, C. Kollath, and M. Köhl, *Higgs mode in a strongly interacting fermionic superfluid,* Nat. Phys. **14**, 781 (2018).

[19]  M. Endres, T. Fukuhara, D. Pekker, M. Cheneau, P. Schauβ, C. Gross, E. Demler, S. Kuhr, and I. Bloch, *The 'Higgs' amplitude mode at the two-dimensional superfluid/Mott insulator transition,* Nature **487**, 454 (2012).

[20]  J. Léonard, A. Morales, P. Zupancic, T. Donner, and T. Esslinger, *Monitoring and manipulating Higgs and Goldstone modes in a supersolid quantum gas,* Science (80-. ). **358**, 1415 (2017).

[21]  N. Bogoliubov, *On the theory of superfluidity,* Ann. J. Phys. **13**, 23 (1947).

[22]  R. Lopes, C. Eigen, N. Navon, D. Clément, R. P. Smith, and Z. Hadzibabic, *Quantum Depletion of a Homogeneous*

*Bose-Einstein Condensate,* Phys. Rev. Lett. **119**, 190404 (2017).

[23] M. Pieczarka, E. Estrecho, M. Boozarjmehr, M. Steger, K. West, L. N. Pfeiffer, D. W. Snoke, A. G. Truscott, and E. A. Ostrovskaya, *Observation of quantum depletion in a nonequilibrium exciton--polariton condensate,* ArXiv 1 (2019).

[24] T. Horikiri, T. Byrnes, and N. Ishida, *Direct photoluminescence observation of the negative Bogoliubov branch in an exciton-polariton condensate,* Quantum Electron. Laser Sci. Conf. QM1G.5 (2012).

[25] T. Horikiri, T. Byrnes, K. Kusudo, N. Ishida, Y. Matsuo, Y. Shikano, A. Löffler, S. Höfling, A. Forchel, and Y. Yamamoto, *Highly excited exciton-polariton condensates,* Phys. Rev. B **95**, 245122 (2017).

[26] J. Kasprzak, M. Richard, S. Kundermann, A. Baas, P. Jeambrun, J. M. J. Keeling, F. M. Marchetti, M. H. Szymańska, R. André, J. L. Staehli, V. Savona, P. B. Littlewood, B. Deveaud, and L. S. Dang, *Bose–Einstein condensation of exciton polaritons,* Nature **443**, 409 (2006).

[27] R. Balili, V. Hartwell, D. Snoke, L. Pfeiffer, and K. West, *Bose-Einstein condensation of microcavity polaritons in a trap,* Science (80-. ). **316**, 1007 (2007).

[28] R. T. Brierley, P. B. Littlewood, and P. R. Eastham, *Amplitude-mode dynamics of polariton condensates,* Phys. Rev. Lett. **107**, 20 (2011).

[29] M. Steger, C. Gautham, D. W. Snoke, L. Pfeiffer, and K. West, *Slow reflection and two-photon generation of microcavity exciton–polaritons,* Optica **2**, 1 (2015).

[30] Y. Sun, P. Wen, Y. Yoon, G. Liu, M. Steger, L. N. Pfeiffer, K. West, D. W. Snoke, and K. A. Nelson, *Bose-Einstein Condensation of Long-Lifetime Polaritons in Thermal Equilibrium,* Phys. Rev. Lett. **118**, 016602 (2017).

[31] Y. Sun, Y. Yoon, S. Khan, L. Ge, M. Steger, L. N. Pfeiffer, K. West, H. E. Türeci, D. W. Snoke, and K. A. Nelson, *Stable switching among high-order modes in polariton condensates,* Phys. Rev. B **97**, 045303 (2018).

[32] V. Kohnle, Y. Léger, M. Wouters, M. Richard, M. T. Portella-Oberli, and B. Deveaud, *Four-wave mixing excitations in a dissipative polariton quantum fluid,* Phys. Rev. B **86**, 1 (2012).

[33] T. Byrnes, T. Horikiri, N. Ishida, M. Fraser, and Y. Yamamoto, *Negative Bogoliubov dispersion in exciton-polariton condensates,* Phys. Rev. B **85**, 1 (2012).

[34] C. J. Pethick and H. Smith, *Chapter 5,* in *Bose-Einstein Condens. Dilute Gasses* (Cambridge University Press, Cambridge, U.K., 2002).

[35] N. S. Voronova, A. A. Elistratov, and Y. E. Lozovik, *Inverted pendulum state of a polariton Rabi oscillator,* Phys. Rev. B **94**, 045413 (2016).

[36] R. Hanai, A. Edelman, Y. Ohashi, and P. B. Littlewood, *Non-Hermitian Phase Transition from a Polariton Bose-Einstein Condensate to a Photon Laser,* Phys. Rev. Lett. **122**, 185301 (2019).

[37] R. Hanai, P. B. Littlewood, and Y. Ohashi, *Photoluminescence and gain/absorption spectra of a driven-dissipative electron-hole-photon condensate,* Phys. Rev. B **97**, 245302 (2018).

[38] M. Gurioli, J. Martinez-Pastor, M. Colocci, A. Bosacchi, S. Franchi, and L. C. Andreani, *Well-width and aluminum-*

*concentration dependence of the exciton binding energies in GaAs/AlxGa1-xAs quantum wells,* Phys. Rev. B **47**, 15755 (1993).

[39] R. Greene, K. Bajaj, and D. Phelps, *Energy levels of Wannier excitons in GaAs-Ga_ {1-x} Al_ {x} As quantum-well structures,* Phys. Rev. B **29**, 1807 (1984).

[40] R. J. Elliott, *Theory of Excitons,* in *Polarons Excit.*, edited by G. D. Kuper, C.G. and Whitfield (Oliver and Boyd, Edinburgh, 1963), pp. 269–293.

[41] M. J. Dugan, H. Georgi, and D. B. Kaplan, *Anatomy of a composite Higgs model,* Nucl. Physics, Sect. B **254**, 299 (1985).

[42] R. Contino, *Tasi 2009 lectures: The Higgs as a Composite Nambu-Goldstone Boson,* ArXiv **1005.4269**, 1 (2010).

[43] M. Carena, L. Da Rold, and E. Pontón, *Minimal composite Higgs models at the LHC,* J. High Energy Phys. **2014**, (2014).

[44] G. Ferretti and D. Karateev, *Fermionic UV completions of composite Higgs models,* J. High Energy Phys. **2014**, (2014).

[45] M. Wouters and I. Carusotto, *Probing the excitation spectrum of polariton condensates,* Phys. Rev. B **79**, 125311 (2009).

[46] R. Jayaprakash, F. G. Kalaitzakis, G. Christmann, K. Tsagaraki, M. Hocevar, B. Gayral, E. Monroy, and N. T. Pelekanos, *Ultra-low threshold polariton lasing at room temperature in a GaN membrane microcavity with a zero-dimensional trap,* Sci. Rep. **7**, 1 (2017).

[47] J. M. Zajac, E. Clarke, and W. Langbein, *Suppression of cross-hatched polariton disorder in GaAs/AlAs microcavities by strain compensation,* Appl. Phys. Lett. **101**, 041114 (2012).

[48] D. Snoke, *Solid State Physics: Essential Concepts, 2nd Edition, Solid State Physics: Essential Concepts, 2nd Edition*, 2nd ed. (Cambridge University Press, Cambridge, 2019).